# High-linearity power amplifier based on GaAs HBT

Jiwei Huang [1,*], Yi Zhang [2] and xx [3]

1. College of Physics and Information Engineering, Fuzhou University, Fuzhou, 350108, China ; huangjw@fzu.edu.cn
2. School of Advanced Manufacturing, Fuzhou University, Quanzhou 362251, China; 852203429@fzu.edu.cn
3. Affiliation 3; e-mail@e-mail.com
4. Affiliation 4; e-mail@e-mail.com
* Correspondence: huangjw@fzu.edu.cn ; Tel.: (optional; include country code; if there are multiple corresponding authors, add author initials)

**Abstract:** This paper presents a power amplifier designed for Wi-Fi 6E using the 2μm gallium arsenide (GaAs) heterojunction bipolar transistor (HBT) process. By employing third-order intermodulation signal cancellation, harmonic suppression, and an adaptive biasing scheme, the linearity performance of the circuit is improved. To achieve broadband performance, the power amplifier also incorporates gain distribution and multi-stage LC matching techniques. The measurement results indicate that, under a 5V supply voltage, it can achieve S21 greater than 31dB, ΔG of ±0.723dB, and P1dB of 30.6dB within the 5.125GHz-7.125GHz frequency band. The maximum linear output power, which satisfies AM-AM < 0.2dB and AM-PM < 1°, is 26.5dBm, and the layout area is 2.34mm².

## 1. Introduction

With the rapid advancement of wireless communication, wireless local-area networks (WLANs) have experienced remarkable growth, driven by the introduction of new standards such as IEEE 802.11ax (Wi-Fi 6/6E). The Wi-Fi 6/6E standards expand the operational spectrum into the 6 GHz band and utilize high-order modulation schemes such as 1024-QAM to achieve data rates approaching 10 Gbps [1,2]. These enhancements, however, introduce stringent requirements for RF power amplifiers (PAs), primarily due to significantly increased peak-to-average power ratio (PAPR) and the associated demands for stringent error vector magnitude (EVM) control, typically better than −35 dB under high-order modulation [3-5]. Consequently, there is an urgent need for compact and efficient power amplifiers capable of delivering high linearity and stable performance over the extended Wi-Fi 6E frequency band (5.125–7.125 GHz).

To address these challenges, various approaches have been explored in the literature. CMOS-based designs offer advantages in integration and cost, yet their limited breakdown voltage and higher substrate losses constrain their effectiveness at higher power levels. GaAs heterojunction bipolar transistor (HBT) technology, characterized by excellent linearity, efficiency, and robustness at high power, remains widely used in WLAN front-end modules [6-8]. Despite these benefits, existing GaAs HBT PAs still encounter significant limitations. Most designs focus on narrower bands around 2.4 GHz or 5 GHz, necessitating separate modules or additional circuitry to cover the broader 5–7 GHz range introduced by Wi-Fi 6E. These solutions result in increased complexity, larger chip area, and higher cost [9,10].

Prior broadband solutions employing harmonic-tuned or distributed matching networks have shown promise in expanding operational bandwidth. Nonetheless, they frequently suffer from larger die sizes, increased insertion losses, and insufficient linearity control at the extremes of the operating band [11]. Techniques such as digital predistortion (DPD) can effectively enhance linearity but add system-level complexity and cost [12]. Alternatively, active biasing methods are utilized to improve linearity with minimal size and power penalties; however, standalone bias adjustments often fail to sufficiently control amplitude-to-amplitude (AM-AM) and amplitude-to-phase (AM-PM) distortions required by advanced modulation schemes [13]. Harmonic suppression techniques using LC traps can improve linearity and efficiency, yet their narrowband characteristics limit performance consistency over wide frequency ranges [14].

In this paper, we propose a novel fully integrated GaAs HBT PA specifically tailored for high-performance Wi-Fi 6E applications across the 5.125–7.125 GHz band. The proposed design addresses critical limitations identified in prior art through three major innovations:

First, we introduce an innovative third-order intermodulation distortion (IMD3) cancellation technique implemented through a carefully engineered dual-bias transistor configuration. Unlike existing IMD3 cancellation methods, which typically involve complex transformers or multi-path combiners [15], our approach exploits intrinsic device nonlinearities directly at the transistor level, significantly reducing complexity and improving robustness.

Second, an adaptive biasing scheme based on a refined mirrored current source topology is developed to dynamically stabilize transistor bias conditions under varying temperature and input power levels. This adaptive mechanism ensures consistent linear performance across a wide dynamic operating range, effectively mitigating AM-AM and AM-PM distortions and enabling the amplifier to meet the stringent linearity requirements of Wi-Fi 6E without relying on external linearization components.

Third, a broadband output matching network is meticulously designed using multi-stage LC structures combined with harmonic suppression circuits. By carefully optimizing inductor and capacitor values, the proposed network achieves uniform impedance matching, stable gain, and effective suppression of second and third harmonics across the entire 2 GHz bandwidth. Additionally, employing bond-wire parasitic inductances strategically helps minimize insertion losses and chip area.

Experimental results validate that the proposed GaAs HBT PA achieves state-of-the-art performance. It demonstrates a fractional bandwidth exceeding 30%, a peak small-signal gain above 31 dB, and output power at 1-dB compression (P1dB) of approximately 30.6 dBm across the 5.125–7.125 GHz band. Crucially, under 802.11ax 1024-QAM test conditions (80 MHz bandwidth), the PA achieves a linear output power up to 26.5 dBm with AM-AM distortion below 0.2 dB and AM-PM distortion below 1°, surpassing previously reported designs in terms of both linearity and output power at comparable EVM specifications.

In summary, the principal contributions of this work are:

- A fully integrated IMD3 cancellation technique that simplifies implementation and enhances linearity;
- A temperature- and input-power-adaptive biasing circuit for robust and stable linear operation;
- A broadband harmonic-suppressed output matching network enabling wide operational bandwidth and consistent linear performance.

- These innovations collectively establish a new benchmark for integrated GaAs HBT power amplifiers in WLAN applications, significantly advancing the state-of-the-art and offering a practical and efficient solution to meet the demands of next-generation wireless communication systems.

The remainder of this paper details the proposed design methodology, circuit implementation, measurement results, and comprehensive benchmarking against recent advancements.

## 2. Circuit Structure and Principle

*2.1. Overall circuit structure of power amplifier.*

As shown in Figure 1, the overall circuit structure block diagram includes three-stage amplification circuits, biasing networks, and input, output, and inter-stage matching networks. According to the design requirements for gain, the circuit adopts a three-stage amplification circuit cascade structure. Considering the trade-off between efficiency and linearity, the biasing points are gradually reduced in the three-stage amplification circuits. As the output power increases stage by stage, the operating point of the transistor will be shifted under the influence of the signal, and the linearity of the circuit will also decrease accordingly. The transistors in the power-stage amplification circuit are divided into two groups to achieve third-order intermodulation component cancellation with different biasing. Additionally, second and third harmonic traps are added to the output impedance matching network to reduce harmonic components in the output signal and improve circuit linearity. The first two stages of amplification circuits have relatively low output power and are biased in the light AB class operating region to provide a cleaner input signal for the power-stage amplification circuit.

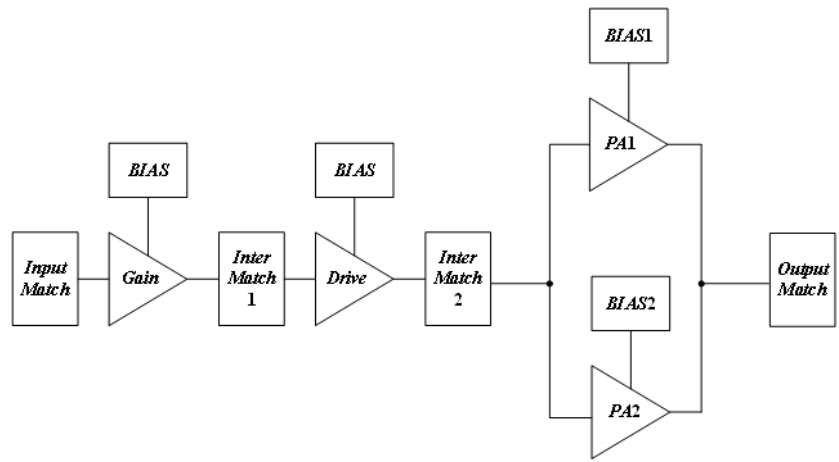

**Figure 1.** Overall circuit structure of power amplifier.

*2.2. Third-Order Intermodulation Component Cancellation Power Stage*

The relationship between each harmonic of the power amplifier and the bias is shown in Figure 2 [16]. It can be seen that as the bias point changes from class A to class C, the amplitude of the fundamental component gradually decreases, and the energy of the harmonic components increases rapidly. Among them, the third-order harmonic component has the most significant impact on the linearity of the power amplifier. The coefficient of the third-order harmonic component is the product of the output impedance and the third-order transconductance [17]. To simplify the analysis, this paper uses the third-order transconductance instead of the third-order harmonic component amplification coefficient for analysis. The fact that the third-order harmonic components have opposite signs under Class AB and Class C biasing indicates that the transistors under Class AB and Class C biasing have opposite signs of third-order transconductance, which provides a theoretical basis for the third-order intermodulation component cancellation scheme.

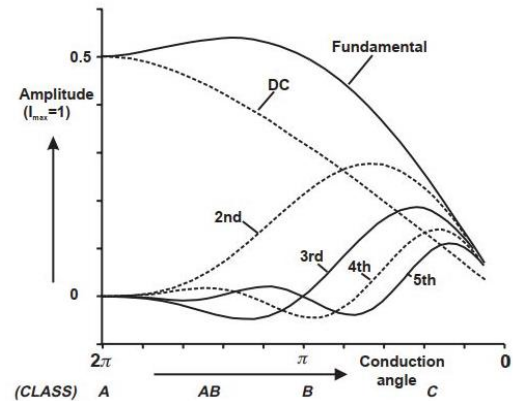

**Figure 2.** Relationship between harmonics and bias of power amplifier [6].

As shown in Figure 3, by adjusting the basic amplification unit of the power - stage amplification circuit, the power transistors are divided into two groups, Bias1 and Bias2, and parallel power combination is carried out with different biases. Thus, the effect that the fundamental-wave signals have the same phase while the third-order intermodulation signals have opposite phases can be achieved. After power combination, the fundamental - wave signals are superimposed and the third - order intermodulation signals are cancelled. Compared with the single - bias - transistor parallel power combination scheme, although the double - bias - transistor parallel power combination scheme for cancelling third - order intermodulation signals sacrifice a certain amount of output power without increasing the circuit complexity, it achieves higher linearity.

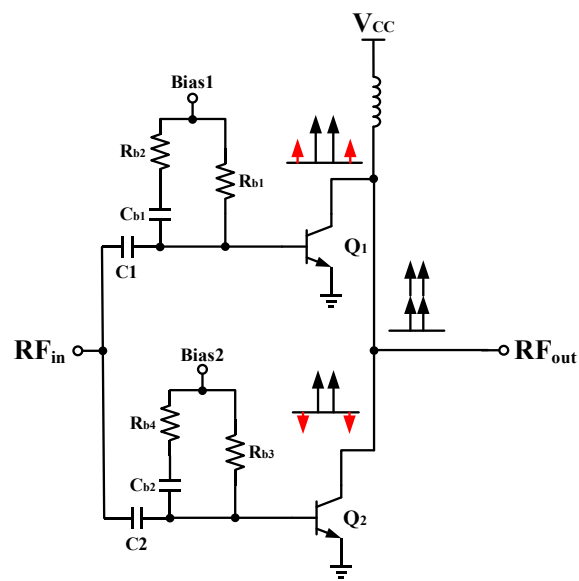

**Figure 3.** Schematic diagram of the third-order intermodulation component offset power stage circuit.

Considering that the parasitic effects of transistors are highly related to transistor biasing, different biasing points can affect the input and output impedances of the two amplification units, thereby influencing the power combination effect and circuit linearity. Therefore, in this paper, the input and output impedances of the two groups of amplification units are simulated respectively. If there is a significant difference, it is necessary to add a power divider or perform parallel power combination after impedance matching for the two groups of power transistors respectively. Use ADS to build the basic unit simulation circuit as shown in Figure 4, and judge the combination scheme according to the simulation results.

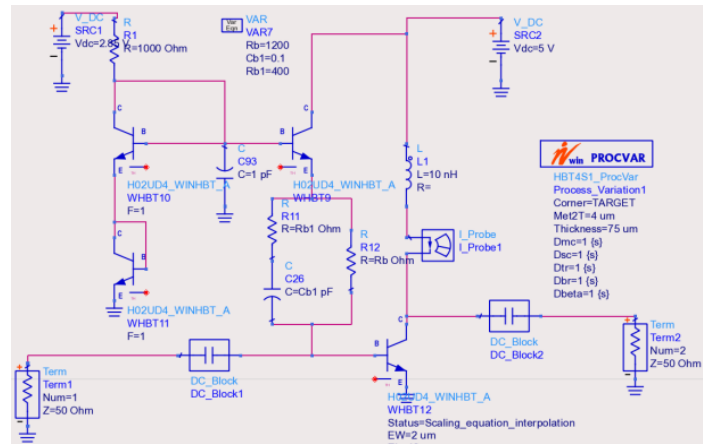

**Figure 4.** Input and output impedance simulation diagram of basic amplifier unit.

As shown in Figure 5, the input and output impedances of two different basic amplification units within the operating frequency band are represented on the Smith chart. It can be observed that under different biases, the input and output impedances of the basic amplification units are affected by the bias but the difference is not significant. Moreover, using power dividers or additional matching circuits within the chip would also occupy extra chip area and cause signal attenuation. Therefore, the design in this paper adopts a direct parallel power synthesis scheme

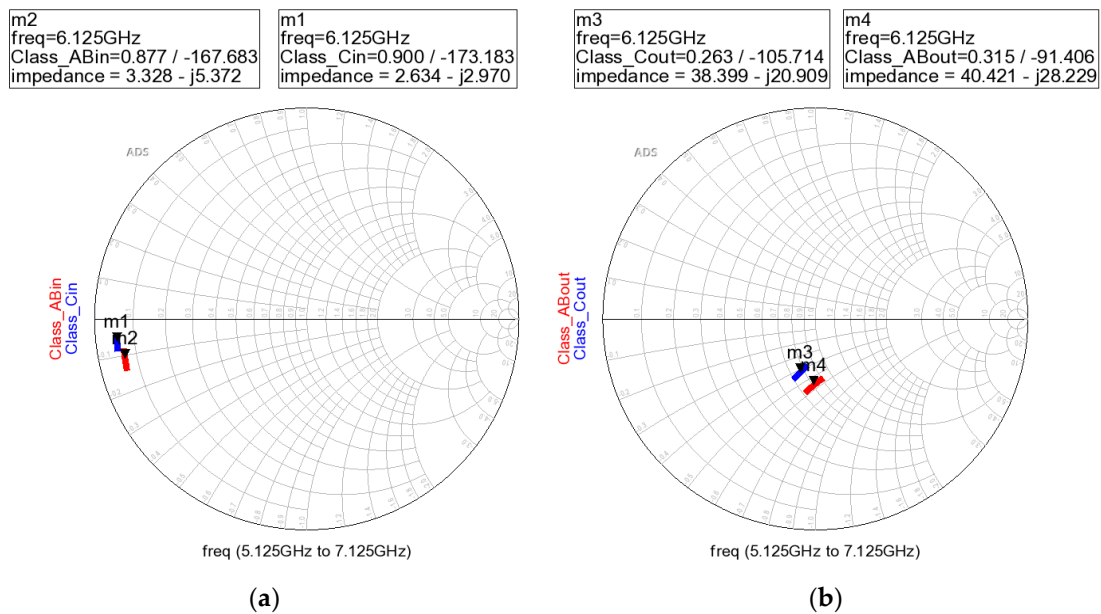

**Figure 5.** Input and output impedances of the operating bands of the two groups of basic amplifier units (**a**) Input impedance of two sets of base amplifier units; (**b**) Output impedance of two sets of base amplifier units.

Using ADS, a dual-tone signal with a frequency spacing of 10MHz at a center frequency of 6.125GHz is input to simulate and observe the current phase and amplitude at the collector stage. The simulation results are shown in Figure 6. Specifically, Figure 6(a) represents the fundamental current phase, where the phase difference between the fundamental signals of the two basic amplification units is within 7°. As can be seen from Figure 6(b), which shows the fundamental current amplitude, the total fundamental current of the two transistors is greater than the branch currents of the two amplification circuits, indicating a superposition relationship. Figure 6(c) illustrates the current phase of the third-order intermodulation signal, indicating that the phase difference of the third-order intermodulation component is approximately 170° when the output power is within the range of 16-27.5dBm. As can be seen from Figure 6(d), which shows the current amplitude of the third-order intermodulation signal, the total output current is smaller than the branch currents of the two amplification circuits within the output power range where there is a cancellation effect, confirming the cancellation effect of the third-order intermodulation component.

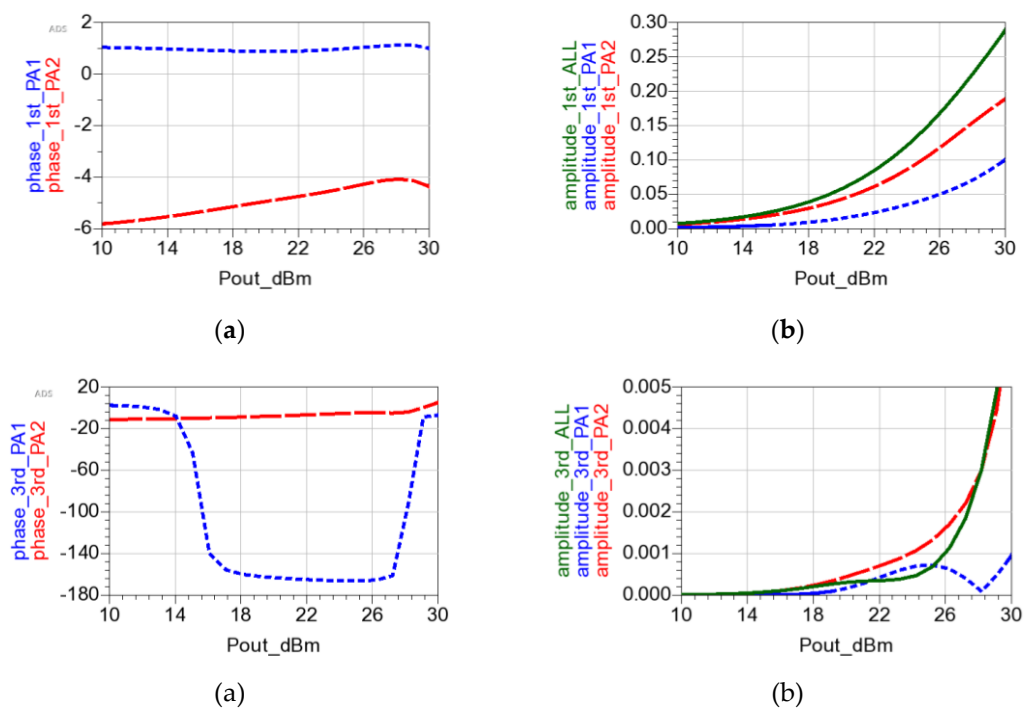

**Figure 6.** Phase and amplitude of the fundamental component of the power level and the third-order intermodulation component: (**a**) Phase of the fundamental component; (**b**) Amplitude of the fundamental component; (**c**) Third-order intermodulation component phase; (**d**) Third-order intermodulation component phase.

For comparison, we adjusted both sets of basic amplification units to AB-class bias and conducted a simulation comparison with the scenario using a third-order intermodulation component cancellation structure. Figure 7(a) shows the simulation results under AB-class bias, while Figure 7(b) shows the simulation results under C-class bias. By comparing the IMD3 metrics, it can be observed that when using the third-order intermodulation component cancellation structure, the IMD3 metrics exhibit a significant decrease within the third-order intermodulation cancellation output power range. This implies that by introducing the third-order intermodulation component cancellation structure, the generation of third-order intermodulation signals can be effectively reduced.

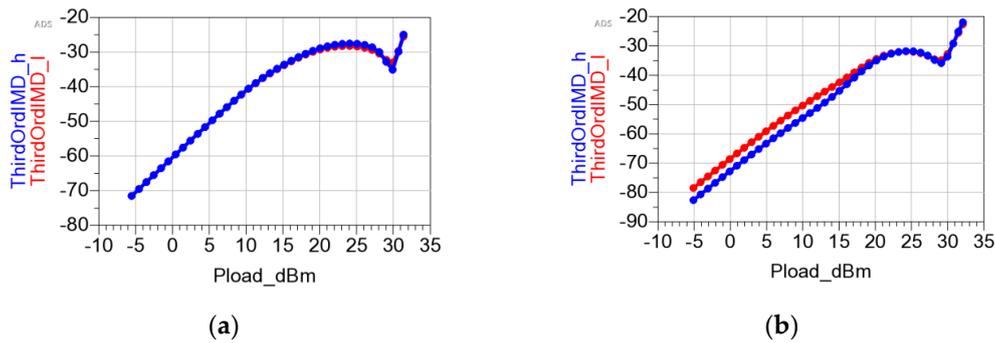

**Figure 7.** Comparison of IMD3 results of Class AB bias and third-order intermodulation component offset bias: (**a**) Single Class AB biased IMD3 curve; (**b**) Third-order intermodulation component cancelling.

*2.3. Adaptive bias circuit*

Based on the analysis of the conditions for third-order intermodulation component cancellation presented above, it is understood that the cancellation scheme requires transistors to have stable bias points; otherwise, it will disrupt the conditions for third-order intermodulation component cancellation. Therefore, a bias circuit that can provide stable bias points plays a crucial role in the effectiveness of third-order intermodulation component cancellation.

Power amplifiers primarily operate in scenarios involving large signal inputs and significant temperature variations. When the input signal increases, due to the diode effect of the base-emitter junction of the HBT transistor, the positive component of the input signal is intercepted. The DC component of this positive signal is superimposed on the static bias current, causing the bias point to shift. Additionally, when the power amplifier operates at high output power for a certain period, the transistor experiences self-heating effects. The increase in temperature will cause Vbe to decrease, altering the bias point position [18]. As shown in Figure 8, under the combined influence of multiple factors, if there is no adaptive bias circuit capable of adjusting the bias, based on the aforementioned analysis, it can be concluded that the bias point will move from A to B. At this point, the transistor's transconductance decreases, reducing circuit gain, deteriorating circuit linearity, and disrupting the conditions for third-order intermodulation component cancellation, thereby narrowing the output power range for cancellation.

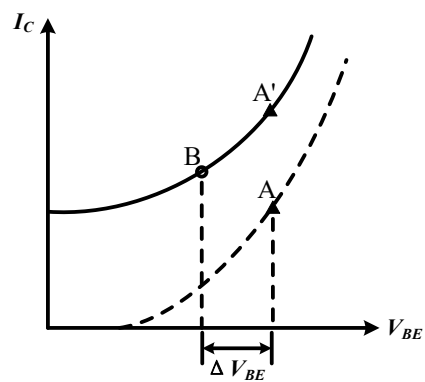

**Figure 8.** Change of power amplifier bias points.

The design in this paper adopts an adaptive bias circuit improved from the mirrored current source, as shown in Figure 9. The ballast resistor Rbias, the active bias transistor Q2, and the linear capacitor Cbias form a radio frequency (RF) signal leakage path, which shorts the RF signals leaked into the bias circuit to ground. The base voltage of Q2 can remain stable under large signal input conditions. The base voltage of the amplifying power transistor Q1 can be expressed as:

$$V_{be1} = V_{b2} - V_{be2} - I_{bias}R_{bias}, \tag{1}$$

As can be seen from equation (1), transistor Q2 can compensate for the voltage drop of transistor Q1. When a large signal input and increased temperature cause the Vbe1 of the power transistor to decrease, Vbe2 simultaneously decreases to compensate for the bias voltage of the power transistor, thereby improving linearity.

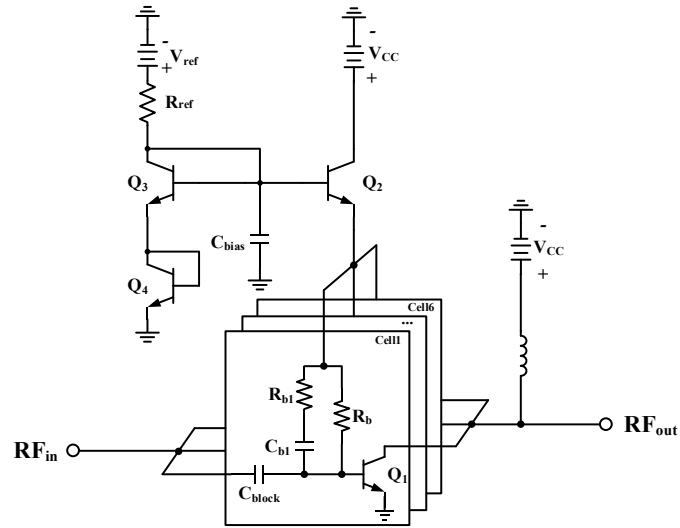

**Figure 9.** Structure of adaptive bias circuit.

To verify the effects of adaptive biasing on temperature compensation and large signal input compensation at the bias point, ADS was used to construct both a simple resistive voltage divider bias circuit as shown in Figure 10(a) and an adaptive bias circuit as shown in Figure 10(b) as the bias simulation circuits for the transistor. During the simulation, temperature variations from -40°C to 160°C and input signal variations from -40dBm to 20dBm were observed separately to study the changes in the transistor's base voltage under different bias circuit conditions.

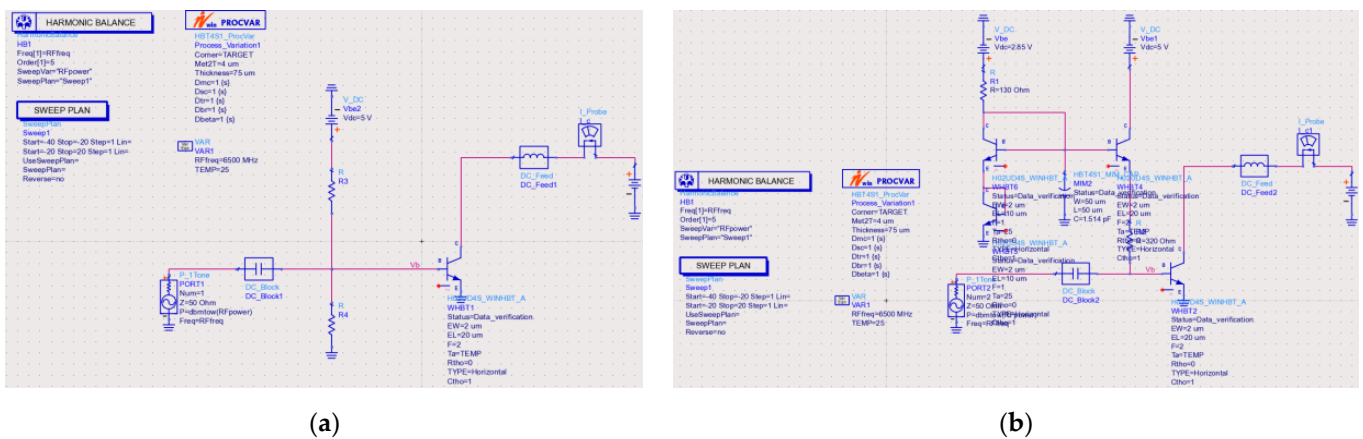

(**a**)          (**b**)

**Figure 10.** Simulation circuit of bias effect on bias point: (**a**) Simple resistance partial bias circuit; (**b**) Adaptive bias circuit.

As shown in Figure 11(a), the base voltage varies with temperature. It can be observed that the temperature variation range of the adaptive bias is smaller compared to the simple resistive voltage divider bias, indicating that the adaptive bias has a certain temperature compensation effect. As shown in Figure 11(b), the base voltage varies with the input signal. It can be seen that compared to the simple resistive voltage divider bias, the adaptive bias exhibits higher tolerance to the decrease in base voltage due to large input signals. Therefore, the adaptive bias has a more significant base voltage compensation capability compared to the simple resistive voltage divider bias.

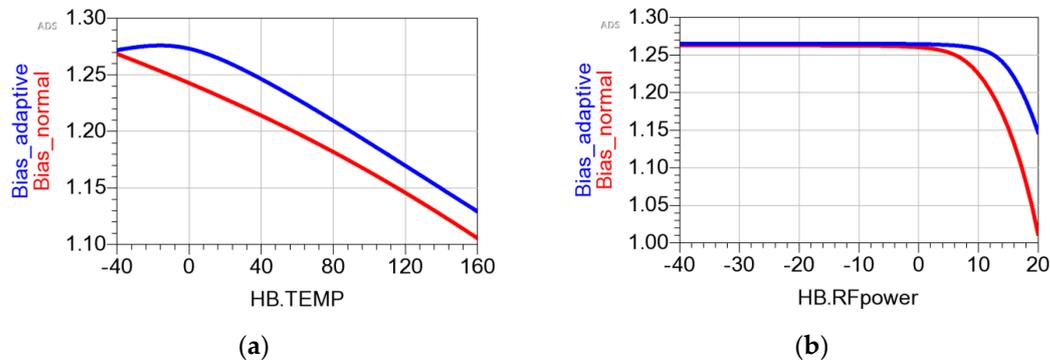

(a)  (b)

**Figure 10.** Simulation results of the influence of bias circuit on bias points: (**a**) The base voltage varies with temperature; (**b**) The base voltage varies with the input signal.

*2.4. Output matching network design*

The schematic diagram of the broadband output impedance matching network with the introduction of harmonic traps in the design of this paper is shown in Figure 12. However, due to the higher resonant frequency, it can easily lead to the problem of excessively small inductance values. In the design of on-chip inductors, lower inductance values result in lower quality factors, which can lead to poor harmonic suppression and increased insertion loss in output impedance matching. To address this issue, in the design of this paper, the series resonant trap is split into two parallel paths to increase the inductance value in the series resonant trap. However, the capacitance value of each path after splitting will also decrease correspondingly. Therefore, when designing the initial matching network, the capacitance value to ground can be consciously increased.

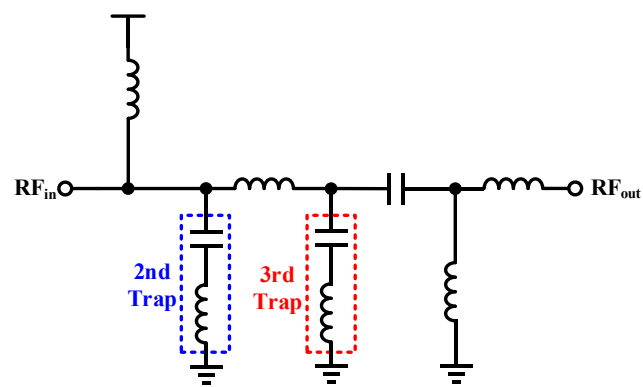

**Figure 12.** Harmonic suppression broadband output impedance matching network.

The simulation results of the S21 curve for the output impedance matching network are shown in Figure 13. It can be observed that the insertion loss of the output impedance matching network is controlled between 0.5-0.7dB within the operating frequency band, and it exhibits good suppression capability for both the second and third harmonics.

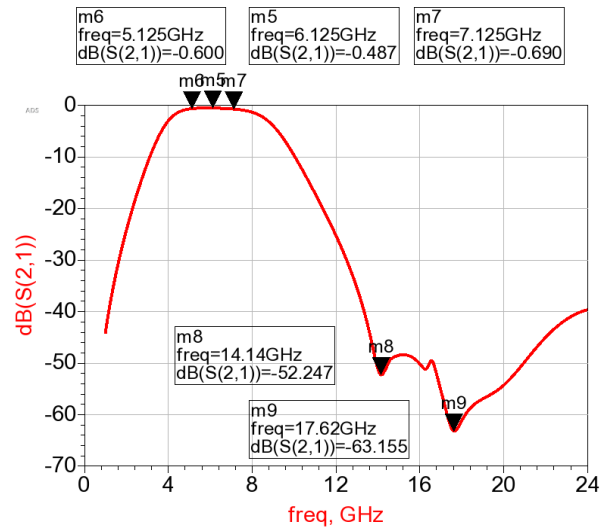

**Figure 13.** Simulation results of insertion loss in wideband harmonic suppression output impedance matching network.

The overall circuit schematic is shown in Figure 14. To enhance the integration level of the chip, both the input impedance matching circuit and the output impedance matching circuit are implemented on the chip, with some inductors utilizing short bonding wire solutions. Considering the impact of parasitic effects from the metal connections in the layout on impedance matching, a concurrent approach of joint simulation and layout iterative design is adopted in the design process. Based on the deviation of joint simulation results from ideal simulation results, continuous optimization and adjustments are made to the layout. During the design process, a sequential approach is followed, starting from the posterior-stage circuits and moving towards the anterior-stage circuits. Ideal connections are gradually replaced with metal connections, ideal devices are substituted with process library devices, and the ideal ground is replaced with ground vias, thereby progressively refining the design of the overall circuit.

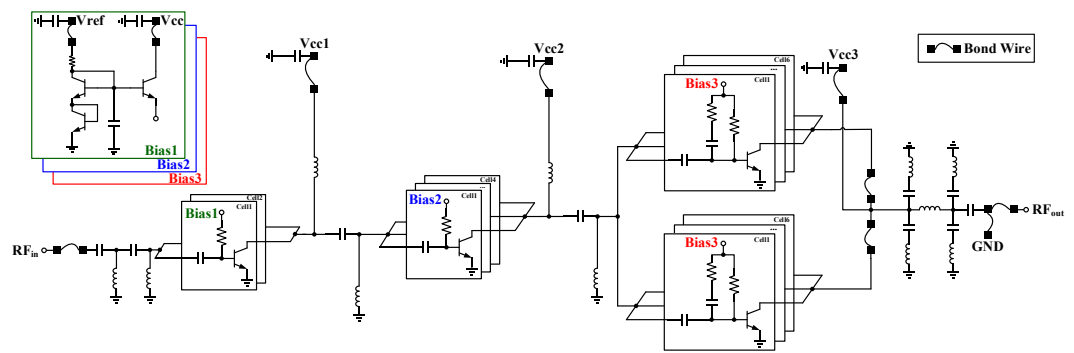

**Figure 14.** Schematic diagram of power amplifier applied to Wi-Fi 6E routing end.

## 3. Results

### 3.1. Overall circuit layout of power amplifier

After the joint simulation results of the layout schematic diagrams of each module circuit meet the preset specifications, the modules are cascaded for electromagnetic simulation, and the entire layout is verified through Design Rule Check (DRC) using Cadence software. Figure 15 shows the layout designed in this paper, using a 2um GaAs HBT process. Both input and output impedance matching are implemented on-chip, and the chip size is 1.71mm*1.37mm.

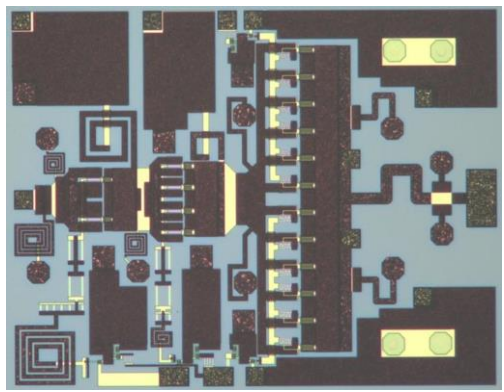

**Figure 15.** Photo of the power amplifier chip.

*3.2. Power amplifier co-simulation results*

After removing the active devices from the layout, the remaining passive components are simulated for electromagnetic parameter extraction to generate a model with parasitic data. This model is then connected with the active devices for joint simulation to obtain simulation results closer to reality. The process library provides three process corners: Fast mode, Target mode, and Slow mode. Initially, the circuit stability under different process corners is simulated. When the stability factor is greater than 1, the circuit is unconditionally stable. The simulation results are shown in Figure 16, indicating that the minimum stability factor within the operating frequency band is 2.269.

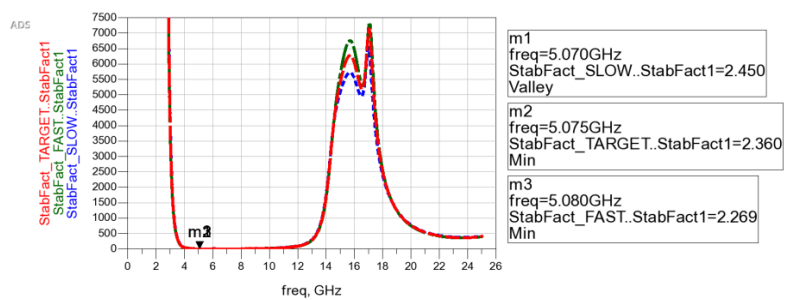

**Figure 16.** Stability coefficient of power amplifier at different process angles.

The S21 gain simulation results are shown in Figure 17(a). The gain is greater than 32dB across the 5.125-7.125GHz frequency band, with a gain flatness variation of less than 1dB, meeting the design specification requirements. The S11 input reflection coefficient simulation results are shown in Figure 17(b). Within the band, it is less than <-12dB, indicating good matching performance. The S22 output reflection coefficient simulation results are shown in Figure 17(c). Within the band, it is <-9dB, meeting the specification requirements.

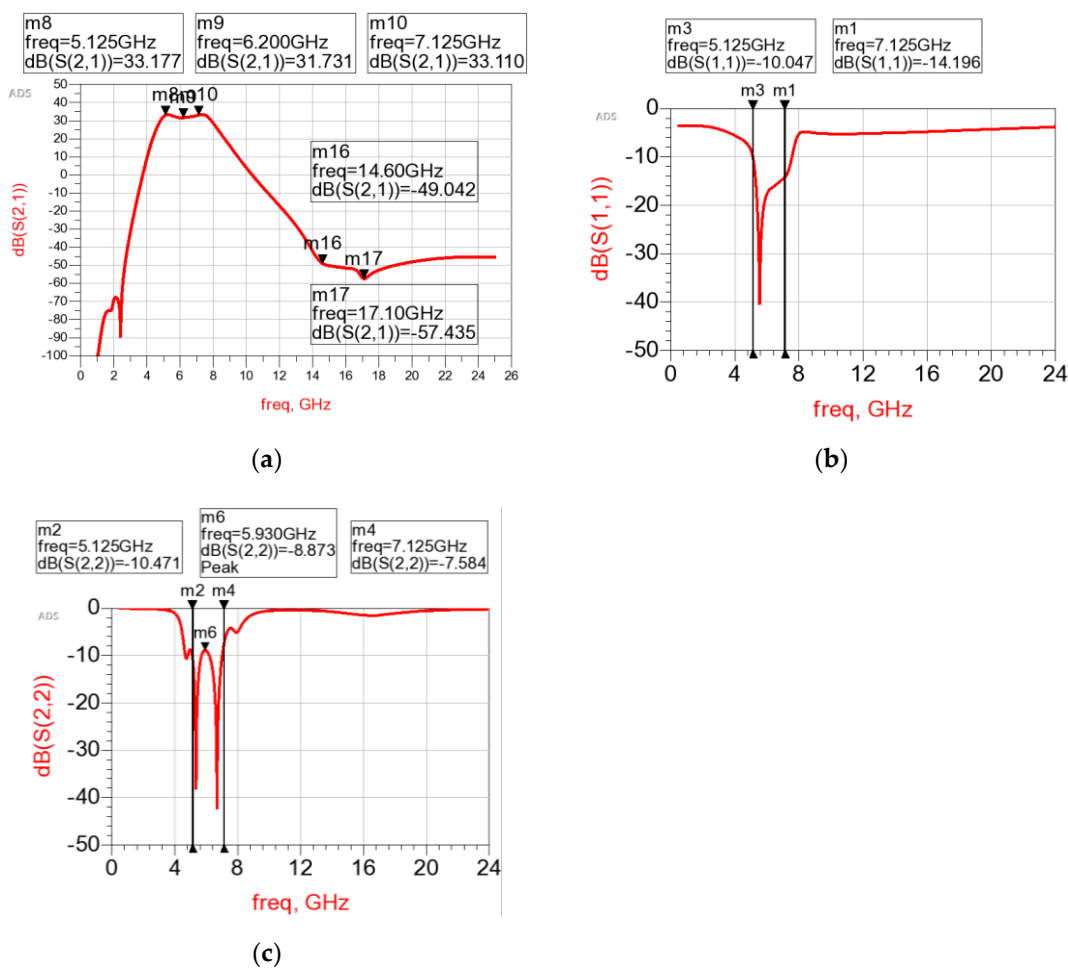

**Figure 17.** Small signal S-parameter co-simulation results: (**a**) S21 co-simulation results; (**b**) Input reflection coefficient; (**c**) Output reflection coefficient.

Figure 18(a) shows the gain curves at frequencies of 5.125GHz, 6.125GHz, and 7.125GHz. When the output power is less than 26.5dBm, the AM-AM distortion is less than 0.1dB. It can be observed that the P1dB at these three frequency points is greater than 30dB, meeting the design specification requirements. Figure 18(b) presents the output phase curve, indicating that the AM-PM distortion is within 1°, satisfying the linearity specification requirements. Figure 18(c) shows the PAE (Power-Added Efficiency) curve, where the PAE at these three frequency points is greater than 26% when the output power is above 30dBm.

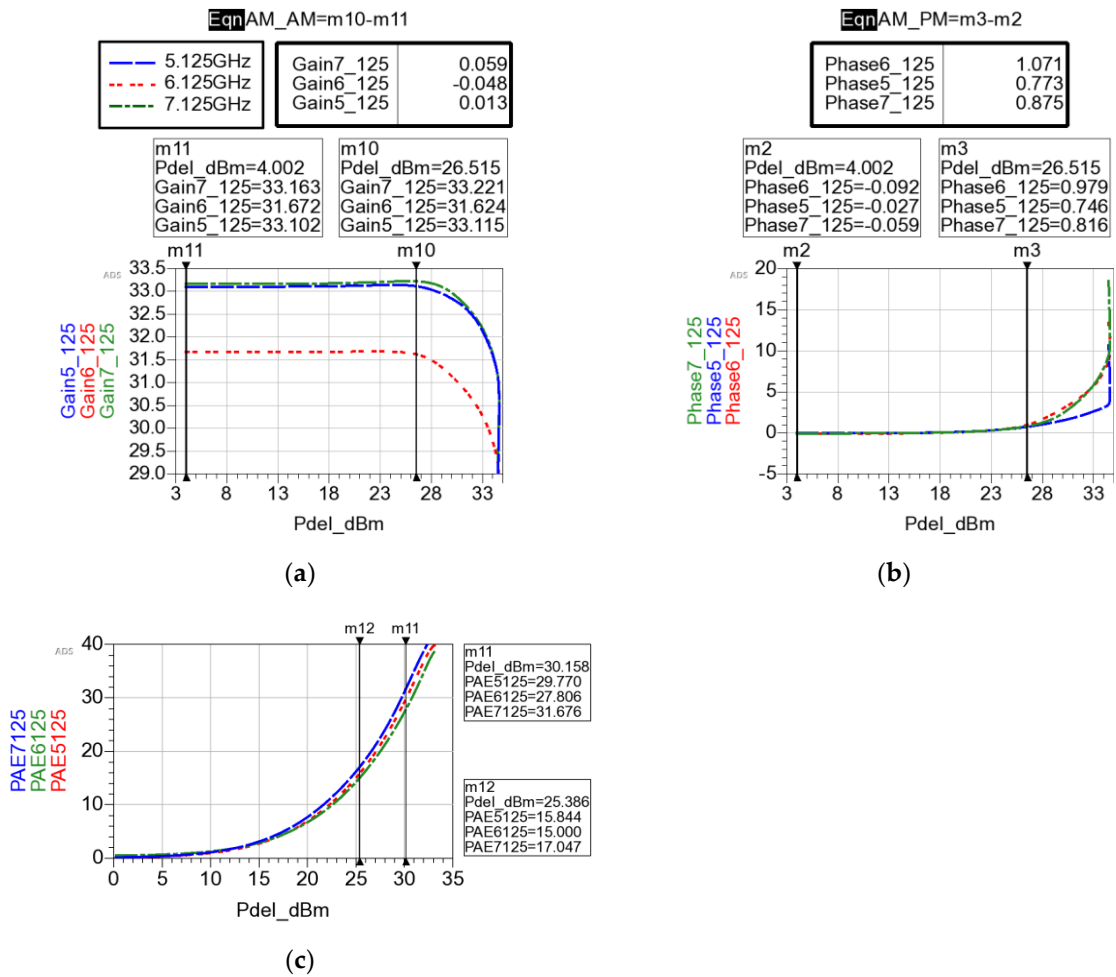

**Figure 18.** Co-simulation curves of AM-AM, AM-PM and PAE: (**a**) Gain change curve with output signal; (**b**) Phase change curve with output signal; (**c**) PAE co-simulation results.

Table 1 summarizes the simulation results of the main performance indicators of the designed Wi-Fi 6E power amplifier, comparing the performance indicators and chip sizes of two power amplifiers with the test results of GaAs HBT linear power amplifiers applied in the 5GHz and 6GHz frequency bands published in international authoritative journals in recent years. In the simulation stage, the linear performance of the power amplifier is mainly characterized by AM-AM and AM-PM distortion. Through comparison, it can be found that compared with reference [19], the design in this paper has a larger bandwidth and a smaller area while achieving higher gain, under the premise of slightly better linearity. Compared with reference [20], although the design in this paper has a larger area, it performs better in linearity and has less variation in gain flatness

**Table 1.** Comparison of power amplifier indexes.

| specification | This work | 错误!未找到引用源。 JSSC'20 | 错误!未找到引用源。 TMTT'23 |
|---|---|---|---|
| Technology | 2μm GaAs HBT | 1μm GaAs HBT | 2μm GaAs HBT |
| Freq(GHz) | 5.125~7.125 | 5~6 | 5.15~7.125 |
| Supply(V) | 5 | 5 | 5.5 |
| Gain(dB) | 31.731 | 6.2 | 24 |
| ΔG(dB) | ±0.723 | ±0.6 | ±1.5 |
| |S11|(dB) | 18 | 15 | 10 |
| |S22|(dB) | 8 | 20 | 5 |
| P1dB(dBm) | 30.6 | 30 | 32.5 |
| PAE*(%) | 27.8 | 25 | 25 |
| Pout**(dBm) | 26.5 | 25 | 22 |
| Area(mm$^2$) | 2.34 | 2.85 | 1.56 |

[1] PAE* is the power additional efficiency when the output power is equal to 30dBm.

## 4. Conclusion

This paper presents a high-power Wi-Fi 6E broadband high-linearity power amplifier designed using the 2μm InGaP/GaAs HBT process. Employing monolithic microwave integrated circuit design technology, the output matching network is integrated within the chip. A constant-impedance broadband harmonic rejection output matching network is designed to ensure the linearity, bandwidth, and other performance metrics of the power amplifier. From the perspective of balancing gain, linearity, bandwidth, and other indicators, an inter-stage matching network with broadband equal-power transmission characteristics is designed to improve the linear performance of the power amplifier across the entire operating frequency band. A theoretical analysis and effectiveness verification of the third-order intermodulation component cancellation scheme used in the power-stage amplification circuit are conducted, which enhances the circuit linearity without increasing circuit complexity. Simulation results indicate that, powered by a 5V supply voltage, the designed Wi-Fi 6E broadband high-linearity power amplifier achieves a small-signal gain of over 32dB within the operating frequency band, with P1dB stable within the range of 30dBm to 31dBm. It can meet the linear specifications for signal amplitude and phase output of the 802.11 ax technical standard, delivering a maximum linear power of 26.5dBm within the frequency range of 5.125GHz to 7.125GHz. The core chip area is 2.34mm2, effectively satisfying the application requirements of high-performance Wi-Fi 6E linear power amplifiers.

preparation, X.X.; writing—review and editing, X.X.; visualization, X.X.; supervision, X.X.; project administration, X.X.; funding acquisition, Y.Y. All authors have read and agreed to the published version of the manuscript." Please turn to the [CRediT taxonomy](CRediT taxonomy) for the term explanation. Authorship must be limited to those who have contributed substantially to the work reported.

**Funding:** Please add: "This research received no external funding" or "This research was funded by NAME OF FUNDER, grant number XXX" and "The APC was funded by XXX". Check carefully that the details given are accurate and use the standard spelling of funding agency names at https://search.crossref.org/funding. Any errors may affect your future funding.

**Data Availability Statement:** We encourage all authors of articles published in MDPI journals to share their research data. In this section, please provide details regarding where data supporting reported results can be found, including links to publicly archived datasets analyzed or generated during the study. Where no new data were created, or where data is unavailable due to privacy or ethical restrictions, a statement is still required. Suggested Data Availability Statements are available in section "MDPI Research Data Policies" at https://www.mdpi.com/ethics.

**Acknowledgments:** In this section, you can acknowledge any support given which is not covered by the author contribution or funding sections. This may include administrative and technical support, or donations in kind (e.g., materials used for experiments). Where GenAI has been used for purposes such as generating text, data, or graphics, or for study design, data collection, analysis, or interpretation of data, please add "During the preparation of this manuscript/study, the author(s) used [tool name, version information] for the purposes of [description of use]. The authors have reviewed and edited the output and take full responsibility for the content of this publication."

**Conflicts of Interest:** Declare conflicts of interest or state "The authors declare no conflicts of interest." Authors must identify and declare any personal circumstances or interest that may be perceived as inappropriately influencing the representation or interpretation of reported research results. Any role of the funders in the design of the study; in the collection, analyses or interpretation of data; in the writing of the manuscript; or in the decision to publish the results must be declared in this section. If there is no role, please state "The funders had no role in the design of the study; in the collection, analyses, or interpretation of data; in the writing of the manuscript; or in the decision to publish the results".